\documentclass{article}
\usepackage{sbctemplate}
\usepackage{adjustbox}
\usepackage{textcomp}
\usepackage{amsmath}
\usepackage[utf8]{inputenc}  
\usepackage{makecell}
\usepackage{graphics}
\usepackage{caption2}

\usepackage{footnote}
\usepackage{tablefootnote}
\usepackage{environ}
\usepackage{xurl}
\usepackage{algpseudocode}
\usepackage{amssymb}
\usepackage[english]{babel}
\usepackage{multirow}
\usepackage{adjustbox}
\usepackage{subfigure}
\usepackage{booktabs}
\usepackage{hyperref}
\usepackage{lscape}
\usepackage{graphicx}
\usepackage{booktabs}
\usepackage{amssymb}

\usepackage[table]{xcolor}

\definecolor{lightgray}{gray}{0.9}
\usepackage{algorithm2e}

\NewEnviron{NORMAL}{%
    \scalebox{2}{$\BODY$} 
} 
 
\NewEnviron{SHORT}{%
    \scalebox{1}{$\BODY$} 
}

\title{{Residual-based Adaptive Huber Loss (RAHL) - Design of an improved Huber loss for CQI prediction in 5G networks}}

\author{Mina Kaviani\inst{1}, Jurandy Almeida\inst{1}, Fábio L. Verdi\inst{1}}

\address{Department of Computer Science (Dcomp) -- Federal University of São Carlos (UFSCar)\\
  Sorocaba -- SP -- Brazil.
    \email{mina.kaviani@estudante.ufscar.br,jurandy.almeida@ufscar.br,verdi@ufscar.br}
}

\begin{document}

\maketitle
\begin{abstract}
The Channel Quality Indicator (CQI) plays a pivotal role in 5G networks, optimizing infrastructure dynamically to ensure high Quality of Service (QoS). Recent research has focused on improving CQI estimation in 5G networks using machine learning. In this field, the selection of the proper loss function is critical for training an accurate model.
Two commonly used loss functions are Mean Squared Error (MSE) and Mean Absolute Error (MAE). Roughly speaking, MSE put more weight on outliers, MAE on the majority. Here, we argue that the Huber loss function is more suitable for CQI prediction, since it combines the benefits of both MSE and MAE.
To achieve this, the Huber loss transitions smoothly between MSE and MAE, controlled by a user-defined hyperparameter called delta. However, finding the right balance between sensitivity to small errors (MAE) and robustness to outliers (MSE) by manually choosing the optimal delta is challenging.
To address this issue, we propose a novel loss function, named Residual-based Adaptive Huber Loss (RAHL). In RAHL, a learnable residual is added to the delta, enabling the model to adapt based on the distribution of errors in the data. Our approach effectively balances model robustness against outliers while preserving inlier data precision.  The widely recognized Long Short-Term Memory (LSTM) model is employed in conjunction with RAHL, showcasing significantly improved results compared to the aforementioned loss functions. The obtained results affirm the superiority of RAHL, offering a promising avenue for enhanced CQI prediction in 5G networks.
%
%
%
\end{abstract}
     
\section{Introduction} \label{sec:Introduction}
The deployment of Fifth Generation Networks (5G) represents a significant leap forward in telecommunications, introducing a trio of services designed to enable enhanced mobile broadband (eMBB), ultra-reliable low-latency communications (uRLLC), and massive machine-type communications (mMTC). The efficacy of these networks hinges on the proficient management of the 5G Core, where Network Functions (NF) play a pivotal role. To ensure optimal communication, radio signal quality indicators are indispensable in managing 5G links. These indicators, including CQI, Signal-to-Noise Ratio (SNR), Reference Signal Received Quality (RSRQ), Reference Signal Received Power (RSRP), and Received Signal Strength Indicator (RSSI), offer valuable insights into communication link quality. User equipment (UE) collects these indicators and communicates them to the evolved Node B (eNB), which serves as the base station. The eNB's radio network controller adjusts channel modulation based on this information to enhance communication links for UEs (User Equipment). 

However, the reactive nature of this process poses challenges, as the collected indicators reflect events in the recent past, and relying solely on reactive operations may not suffice for optimal performance, especially with 5G links characterized by short-range, high-frequency radio signals and mobile UEs. To overcome this limitation and foster more proactive network management, there is a growing need for predictive analytics and machine learning algorithms. These technologies can analyze historical data, identify patterns, and forecast potential issues or changes in the network. Leveraging predictive insights empowers network operators to take proactive measures, optimizing channel modulations, resource allocation, and overall network performance \cite{yin2020predicting, parera2019transfer, vankayala2020neural, sakib2020deep, kimura2021deep}. 

In the design of the communication system, the CQI is a crucial parameter in communication systems. It encodes the state of the channel, allowing base stations to adjust service quality based on real-time channel conditions. This facilitates efficient communications \cite{yin2020predicting}. However, accurately forecasting CQI proves challenging due to its dynamic nature, ranging from 0 to 15 and influenced by various environmental factors. Incorrect predictions can significantly degrade the 5G channel's quality, impacting modulation and resource allocation by the base station, which relies on reported CQI to optimize bandwidth usage. This can lead to decreased Quality of Experience (QoE) for users, affecting application data rates and wasting network resources.

Machine learning models, though powerful, face difficulties in accurately predicting CQI due to abrupt shifts and fluctuations, potentially leading to suboptimal performance. As a result, alternative methodologies are necessary. Determining the appropriate error metric for evaluating CQI signal quality, whether Mean Squared Error (MSE) or Mean Absolute Error (MAE), is complicated by the limitations of both metrics and the specific conditions affecting CQI accuracy within the 5G ecosystem.

To address this, we explore the Huber loss function, known for its robust, piece-wise structure that mitigates the influence of outliers compared to MSE \cite{khan2016application}, \cite{raca2020leveraging}. Huber loss function \cite{gokcesu2021generalized} seamlessly blends the quadratic (MSE) and absolute value (MAE) losses, offering a user-controlled trade-off via 
a hyperparameter called delta. However, manually setting this hyperparameter to balance sensitivity to small errors (MAE) and robustness to outliers (MSE) can be challenging. Motivated by this, instead of manually setting this hyperparameter, which is hard, we propose the Residual-based Adaptive Huber Loss (RAHL), which transforms the hyperparameter delta into a trainable parameter. By transforming delta into a trainable parameter, RAHL empowers models to learn optimal outlier robustness during training, achieving a sweet spot between outlier resistance and inlier accuracy~\cite{dong2020training,gokcesu2021generalized}.

Through a comprehensive investigation and together with the LSTM (Long Short-Term Memory) model for CQI prediction, we systematically evaluated the impact of RAHL in model training, comparing its performance to alternative loss functions like the ``standard'' Huber, MSE and MAE. Employing Mean Absolute Percentage Error (MAPE) as the evaluation metric, our results revealed that the RAHL consistently produced lower MAPE values compared to other loss functions, indicating improved model accuracy. This research contributes to a deeper understanding and broader application of machine learning models in forecasting signal quality indicators, ultimately leading to enhanced 5G network performance.

This paper is organized as follows: Section \ref{sec:Related Works} delves into related works representing the current state of the art, Section \ref{sec:System model} outlines the characteristics and operational details of the proposed LSTM model and RAHL, Section \ref{sec:Evaluation and results} presents the main quantitative results across various scenarios, and Section \ref{sec:Conclusion} offers conclusions and recommendations for future research endeavors.

\section{Related works} \label{sec:Related Works}
Researchers in anticipatory networking, where accurate prediction of wireless channel quality is crucial, have traditionally used past channel measurements to guide future forecasts. However, a recent study by \cite{parera2019transfer} boldly tackles the challenge of cross-channel quality prediction. Their innovative transfer learning framework harnesses the combined power of CNNs and LSTMs to forecast channel quality for specific frequency carriers. Notably, their work employs two distinct model architectures, each trained with the RMSE loss function, and reveals LSTM's superior performance among various evaluated algorithms.

Numerous challenges in the fields of learning, optimization, and statistics literature \cite{cesa2006prediction, portnoy2000robust} underscore the need for resilient solutions, mandating that models undergo training or optimization with diminished susceptibility to outliers. This ensures their robustness against outlier influence, in contrast to the impact of inliers, i.e., nominal data \cite{hastie2015statistical, huber2004robust}. This approach finds widespread application in tasks related to parameter estimation and learning, particularly in cases where prioritizing a robust loss, such as the absolute error, proves more advantageous than opting for a non-robust loss like the quadratic error due to its resistance to substantial errors. When faced with the choice, it becomes essential to transcend traditional outlier detection techniques \cite{gokcesu2018sequential, delibalta2016online}, and focus efforts on incorporating inherent resilience to outliers into the design of loss functions. In this context, the Huber loss function emerges as a promising choice, striking a balance between the mean squared error and absolute error, offering robustness to outliers while maintaining sensitivity to inliers. Its adaptive nature makes it well-suited for scenarios where a compromise between the two extremes is crucial for model performance and generalization \cite{gokcesu2021generalized}.

Adaptive Huber regression is a robust and data-driven solution for handling outliers and heavy-tailed distributions in big data, unlike traditional methods. It automatically adjusts parameters to balance bias and robustness, proven effective across various data scenarios, including those with heavy-tailed distributions \cite{sun2020adaptive}.

\cite{cavazza2016active} propose a method for exact Huber loss optimization in scalar regression with semi-supervised learning. Their approach incorporates multi-view learning, which leverages information from multiple data perspectives, and manifold regularization. Additionally, they employ a data-driven adaptation of the Huber loss threshold and actively balance the use of labeled data to mitigate the impact of noisy or inconsistent annotations during training. 


Unlike \cite{cavazza2016active} and \cite{sun2020adaptive}, which try to learn the hyperparameter delta itself, RAHL tries to learn a residual that is added to it. In this way, we easy the training of deep models, leading to improved performance.

\section{System model} \label{sec:System model}
In this paper, we introduce RAHL for CQI prediction. Our model strategically leverages temporal dependencies, spectral features, and historical data to enhance its predictive capabilities, aiming for optimal performance in CQI estimation.






\subsection{The deep learning-based CQI prediction model}  \label{subsec:The deep learning-based CQI prediction model}

Building on previous work, we employ the well-known LSTM model due to its ability to capture long-term dependencies in sequential data, which makes it ideal for accurately predicting CQI values. It is not our intention to go deeper in the LSTM architecture but shortly introduce how it works. 

\begin{figure}[!htb]
\footnotesize
\centering
\includegraphics[width=0.55\linewidth]{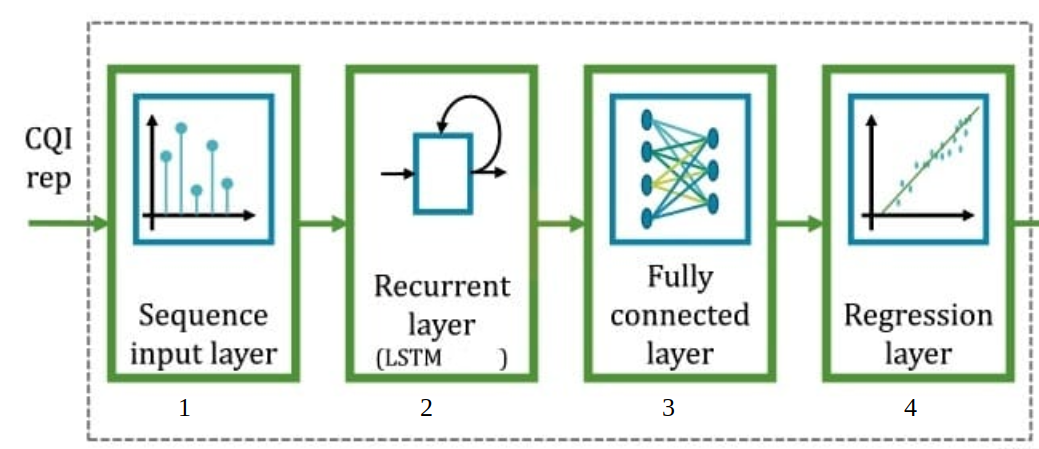}
\caption{Considered LSTM scheme. }
\label{fig:LSTM_scheme}

\end{figure}

Figure \ref{fig:LSTM_scheme} illustrates the LSTM architecture, designed to handle input sequences with one-dimensional features. The sequence first passes through an LSTM layer with 64 hidden units, enabling it to learn sequential patterns. Next, a fully connected layer with 64 units introduces non-linear transformations to the LSTM output. Finally, the regression layer generates a single output value, representing the model's prediction for the given sequence \cite{bartoli2022cqi}. 

\subsection{Residual-based Adaptive Huber Loss (RAHL) - the improved Huber loss for CQI prediction} \label{subsec:Residual-based Adaptive Huber Loss}

In the realm of regression problems, the absolute loss, $L_1(y, f_\theta(x)) = |y - f_\theta(x)|$, and the ubiquitous quadratic (squared) loss, $L_2(y, f_\theta(x)) = (y - f_\theta(x))^2$, emerge as consequential alternatives, requiring a strategic choice rooted in their distinctive characteristics. The quadratic loss, recognized for its robust convexity, facilitates rapid learning rates, while the absolute loss is esteemed for its inherent resilience. This dichotomy emphasizes the need of amalgamating the strengths of both loss functions, leading to models that not only exhibit robustness against outliers but also achieve swift convergence with optimal goodness of fit. 

The Huber loss emerges as a prevalent solution, seamlessly combining quadratic and absolute losses to formulate a resilient loss function that converges quickly~\cite{gokcesu2018sequential}. The Huber loss is widely adopted in regression tasks, especially when dealing with outliers or noise in the data. It achieves a balanced compromise between MSE (quadratic loss) and MAE (absolute loss), offering improved resilience against extreme data points compared to relying solely on MSE or MAE. The pivotal transition point within the Huber loss dictates its shift from quadratic to absolute loss behavior, making it a crucial hyperparameter that significantly influences the performance of a regression model. Nevertheless, the challenge lies in selecting the optimal transition parameter, necessitating frequent hyperparameter searches to identify the most suitable value~\cite{meyer2021alternative}.

Formally, the Huber loss is given by (Equation~\ref{huber}):
\begin{equation}\label{huber}
H(y, f_\theta(x)) =
\begin{cases}
    \frac{1}{2} (y - f_\theta(x))^2, & \text{if } |y - f_\theta(x)| \leq \delta \\
    \delta |y - f_\theta(x)| - \frac{1}{2} \delta^2, & \text{if } |y - f_\theta(x)| > \delta
\end{cases}, 
\end{equation}

\noindent where $y$ is the ground truth, $f_\theta(x)$ is a model defined by the learnable parameters $\theta$, and $\delta$ is a positive hyperparameter that acts as a pivotal regulator, transitioning the penalty from $L_2$ to $L_1$. This crucial hyperparameter balances the critical trade-off between model accuracy and robustness to outliers. Choosing the right value for $\delta$ is crucial, as it determines the transition point where the loss function switches from prioritizing precision to emphasizing robustness. A smaller $\delta$ favors $L_1$ behavior, boosting accuracy but decreasing resilience to outliers. Conversely, a larger $\delta$ pushes the loss function toward $L_2$ characteristics, enhancing robustness but potentially at the cost of accuracy. 

Tuning the hyperparameter $\delta$ by hand poses substantial challenges for adopting the Huber loss to train regression models, which include subjectivity, data dependence, computational burden, potential overfitting, and the difficulty of balancing accuracy and robustness. This requires a cautious approach and exploration of more sophisticated hyperparameter tuning methods for robust regression models. To mitigate these shortcomings, we propose the Residual-based Adaptive Huber Loss (RAHL), a transformative approach that empowers the model to automatically determine the optimal penalty scheme.

\begin{figure}[!htb]
\centering
\includegraphics[width=0.60\linewidth]{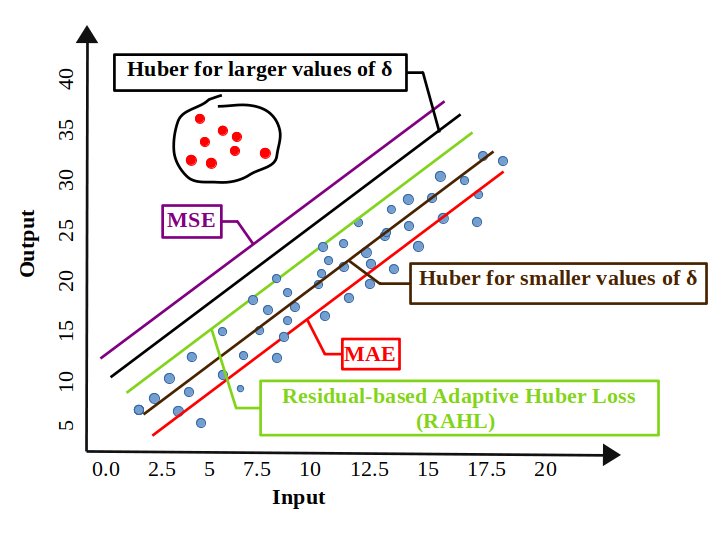}
\caption{Exploring how different loss functions shape regression models.}

\label{fig:MSE-MAE}

\end{figure}

Figure~\ref{fig:MSE-MAE} compares RAHL and the other loss functions, illustrating how outliers affect the solution (i.e., model).
In this figure, outliers are represented by red points and inliers by blue points. Solving regression problems with MSE penalty results in a model (purple line) that heavily leans towards outliers. Conversely, employing MAE yields a model (red line) close to inliers, neglecting outliers (red points). Introducing the Huber loss allows manual tuning of the hyperparameter $\delta$, where a large $\delta$ mimics MSE (black line) and a small $\delta$ mimics MAE (brown line). Consequently, carefully selecting this hyperparameter is vital, as it directly affects the model's sensitivity to outliers and, ultimately, its performance in robust regression tasks. The underlying idea of RAHL is to find a balance, offering sensitivity to small errors (blue points) while maintaining robustness to outliers (red points), as depicted by the green line—ultimately providing the most optimal model for these data.  

Mathematically, RAHL is identical to the Huber loss, but instead of using a fixed value for the hyperparameter $\delta$, it is computed by (Equation~\ref{rahl}):
\begin{align}
RAHL(y, f_\theta(x)) & = 
\begin{cases}
    \frac{1}{2} (y - f_\theta(x))^2, & \text{if } |y - f_\theta(x)| \leq \delta \\
    \delta |y - f_\theta(x)| - \frac{1}{2} \delta^2, & \text{if } |y - f_\theta(x)| > \delta
\end{cases}, \\
\delta & = \alpha + \text{ELU}(\beta), \label{rahl}
\end{align}

\noindent where $\alpha$ is a positive hyperparameter defining the initial value for $\delta$ and $\beta$ is a learnable parameter that is added to $\alpha$. To bound the output and obtain a positive value for $\delta$, the Exponential Linear Unit (ELU)~\cite{clevert2016elu} function is applied to the parameter $\beta$. ELU is an activation function that performs the identity operation on positive inputs and an exponential non-linearity on negative inputs and is given by (Equation~\ref{elu}):
\begin{equation}\label{elu}
ELU(x) = 
\begin{cases}
    x, & \text{if } x \geq 0 \\
    \alpha (e^x - 1), & \text{if } x < 0
\end{cases}, \\
\end{equation}

\noindent where $\alpha$ is a constant that defines function smoothness when inputs are negative and is usually set to 1.0. By setting this constant with the same value chosen for the hyperparameter $\alpha$ from Equation~\ref{rahl} (i.e., the initial value for $\delta$), we constraint its output value to the range $[-\alpha, +\infty)$, ensuring that $\delta$ is always positive.







\begin{figure}[!htb]
\footnotesize
\centering
\includegraphics[width=0.45\linewidth]{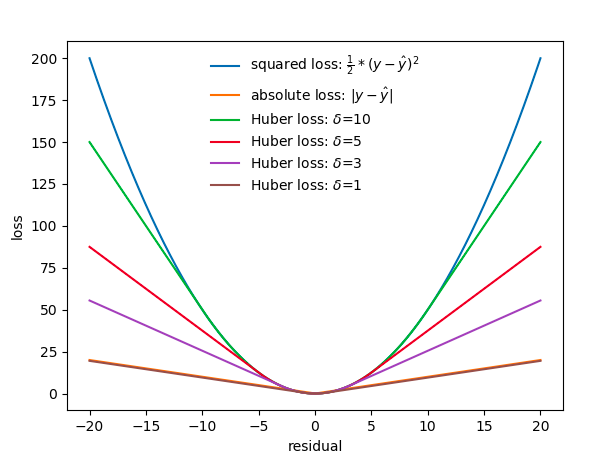}
\caption{The Huber loss for various values of $\delta$, moving between MSE and MAE.}
\label{fig:trad}

\end{figure}

Figure~\ref{fig:trad} illustrates how changing $\delta$ impacts the Huber loss, as compared to absolute and quadratic losses. Each line depicts the loss (y-axis) as a function of the residual (x-axis), which is given by the difference between the groundtruth value ($y$) and the model's prediction ($\hat{y}$).  The larger the $\delta$, the more the Huber loss behaves like MSE, conversely, the smaller the $\delta$, the more the Huber loss behaves like MAE. The key advantage of RAHL is to enable the model to adapt based on the distribution of errors, learning to behave more similar to MSE or MAE based on the data.

\section{Experimental evaluation} \label{sec:Evaluation and results}
This section reveals the intricate details of our data collection and methodology, meticulously outlining each step and ensuring a clear understanding of the research framework.

\subsection{Data collection}
We investigate three datasets:
\begin{itemize}

\item Dataset A:

A dataset comprises Channel Level Metrics (CLM) files and YouTube Quality of Experience (QoE) logs stored in MySQL, featuring metrics like Timestamp, Location, Network details, Signal Strength, Bitrates, Altitude, and Experiment ID (EID). The data, obtained from a comprehensive 5G collection campaign in diverse scenarios (Mobility, Pedestrian, Indoor, Outdoor), utilized the YouTube IFRAME API and Android's Network Monitor app. The dataset captures a wide range of parameters, providing insights into 5G network performance across various use cases~\cite{DATASETA}.

\item Dataset B:

The dataset comprises 83 records of Internet transmissions recorded by G-NetTrack v18.7 on a Samsung S10 connected to an Irish mobile operator. It includes 3142 minutes of transmission logs, organized into three services (File Download, Amazon Prime, and Netflix) and two mobility patterns (Static and Vehicular). The logs, limited by an 80GB data plan, are stored in a CSV file with fixed features and variable data points. The dataset provides detailed attributes such as timestamp, geographical coordinates, node velocity, mobile operator (anonymized), cell ID, network mode, bitrates, device state, and various signal quality indicators for both the primary and neighboring cells~\cite{raca2020beyond}.

\item Dataset C:

Field tests in Brazil through a 5G network using a Samsung S21 5G, focusing on traffic and mobile network monitoring. YouTube metrics captured through various clients were analyzed alongside manual monitoring using G-NetTrack Pro in 5G-covered areas in São Paulo. Data was enriched with Anatel's Mosaico information on registered telecommunication stations, providing details on technologies, equipment, frequencies, locations, licensing, and ownership~\cite{DATASETC}.

\end{itemize}

\subsection{Data preparation}
\begin{figure}[!htb]
\centering
\includegraphics[width=0.55\linewidth]{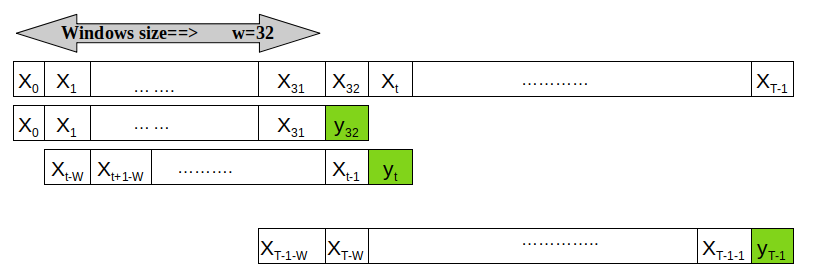}
\caption{Sliding windows for time series forecasting.}

\label{fig:Sliding}

\end{figure}

We tackle missing data challenges by leveraging NaN (Not a Number) for placeholder values, a standard practice in Python's numerical domain. Through pre-processing techniques like \emph{MinMaxScaler}, we achieve uniformity and improve model performance. Recognizing the importance of time order, we employ a sliding window approach (w=32, shifted T times) to preserve temporal context within our analysis, Figure~\ref{fig:Sliding} shows this procedure in detail~\cite{parera2019transfer}.

\subsection{Training procedure}

To leverage the inherent time dependence of our data, we strategically partition it by timestamps, reflecting real-world dynamics where past information heavily influences future predictions. Each CSV dataset is meticulously divided, dedicating 80\% for training and reserving 20\% for rigorous testing.

\subsection{Implementation details}

We implemented the Python code for our project and executed it using Google Colab, a cloud-based platform for collaborative coding and data analysis. Also, we defined the network architecture and training hyperparameters for the LSTM model, which are summarized in Figure~\ref{fig:LSTM_Network_Parameters}.
\begin{figure}[!htb]
    \centering

\adjustbox{max width=\textwidth}{
    
    \begin{tabular}{cp{1cm}c}
        \begin{tabular}{c|c}
            \hline
            Input Size & 1 \\
            Hidden Size & 64 \\
            Number of Stacked Layers & 1 \\
            Number of FC Layers & 1 \\
            FC Hidden Size & 64 \\
            Activation Function for FC Layers & ReLU \\
            Output Size & 1 \\
            \hline
        \end{tabular}
        & &
        \begin{tabular}{c|c}
            \hline
            Number of Epochs & 300 \\
            Mini-Batch Size & 24 \\
            Windows Size & 36 \\
            Initial Learning Rate & 0.01 \\
            Optimizer & Adam \\
            \hline
        \end{tabular} \\
        & & \\
        (a) Network architecture & & (b) Training hyperparameters \\
    \end{tabular}

}

    \caption{Network architecture and training hyperparameters for the LSTM model.}
    \label{fig:LSTM_Network_Parameters}

\end{figure}




\subsection{Performance metrics} 

To ensure a fair comparison of the model performance across different loss functions, we utilize the Mean Absolute Percentage Error (MAPE) . Our approach involves training the model with different loss functions, and using MAPE as the primary validation and testing metric. Widely used in statistics and data analysis, particularly for time series forecasting, MAPE expresses error as a percentage, as shown in Equation~\ref{fig:MAPE}, where lower values indicate better performance~\cite{de2015using}.

\begin{equation}
\label{fig:MAPE}
\text{MAPE} = \frac{1}{n} \sum_{i=1}^{n} \left| \frac{Y_i - \hat{Y}_i}{Y_i} \right| \times 100\%
\end{equation}

\subsection{Experimental results}
\vspace{-0.15cm}

Initially, we trained the LSTM model using the Huber loss and manually selected the hyperparameter $\delta$. To do so, we tested various values for $\delta$, ranging from 0.5 to 4.0 by a step of 0.5. The results obtained for each dataset are presented in Table~\ref{tab:Huber-diffrent-delta} and show the performance variations across diverse values for $\delta$ and distinct categories within datasets, highlighting the lowest MAPE values for each setting. In this table, MAPE can be interpreted as a measure of forecast performance: lower MAPE values indicate more accurate forecasts, while higher MAPE values indicate less accurate forecasts. After extensive computations across various values for $\delta$, the best choice was identified as the minimum error. However, it is crucial to note that this minimum may not be the optimal solution, as further analysis will demonstrate. As expected, there is no silver bullet for all cases: the performance for the Huber loss often depends on the choice for the hyperparameter $\delta$.

\begin{table}[!h]
\centering
\caption{MAPE values for the Huber loss with different values for $\delta$.}
\label{tab:Huber-diffrent-delta}

\adjustbox{max width=\textwidth}{

\begin{tabular}{l|cccccccc}
\textbf{Dataset} & \textbf{$\delta=0.5$} & \textbf{$\delta=1$} & \textbf{$\delta=1.5$} & \textbf{$\delta=2$} & \textbf{$\delta=2.5$} & \textbf{$\delta=3$} & \textbf{$\delta=3.5$} & \textbf{$\delta=4$} \\
\hline
A-Indoor               & 7.55 & 7.56 & 6.89 & 7.42 & 7.72 & 6.33 & 7.55 & 7.10 \\
A-Pedestrian           & 26.05 & 25.79 & 26.81 & 27.66 & 27.34 & 27.57 & 25.83 & 26.38 \\
A-Mobility             & 18.49 & 18.77 & 19.35 & 16.67 & 17.62 & 19.11 & 15.74 & 15.63 \\
A-Outdoor              & 72.24 & 59.05 &  54.13 & 61.22 & 70.31 & 52.74 & 58.13& 56.88 \\
B-Static-Netflix       & 14.61 & 14.29 & 15.46 & 18.05 & 16.35 & 11.93 & 11.19 & 12.87 \\
B-Static-Download      & 18.74 & 13.87 & 15.93 & 14.07 & 13.49 & 16.51 & 16.67 & 15.94 \\
B-Static-Amazon\_Prime  & 1.089 & 1.023 & 1.1 & 1.21 & 1.26 & 1.19 & 1.12 & 1.11 \\
B-Driving-Netflix      & 62.09 & 67.65 & 66.97 & 58.08 & 57.6 & 60.26 & 62.5 & 66.66 \\
B-Driving-Download     & 41.31 & 39.53 & 46.15 & 40.82 & 45.75 & 46.87 & 45.29 & 38.99 \\
B-Driving-Amazon\_Prime & 20.3 & 28.44 & 24.02 & 23.18 & 24.04 &18.81 & 19.61 & 19.08 \\
C                      & 25.59 & 29.25 & 25.48 &23.93 & 23.85 & 23.08 & 22.67 & 24.2 \\
\end{tabular}
}
\end{table}



Table~\ref{tab:Huber-MSE} compares the results for RAHL with those obtained for the Huber loss, considering the best $\delta$ values found  for each setting, according to the previous experiment. Unlike the Huber loss, in RAHL the hyperparameter $\delta$ is transformed into a trainable parameter, eliminating the need for choosing $\delta$ by hand. The results for MSE and MAE were also included for comparison. 
Consistently across all the datasets, the MAPE values were lower when using RAHL, indicating its superior performance relative to the other loss functions examined. 

\begin{table}
\centering
\caption{MAPE values obtained for different loss functions.}
\label{tab:Huber-MSE}

\begin{tabular}{l|cccccccc}
\multirow{2}{*}{\textbf{Dataset}} & \multirow{2}{*}{\textbf{RAHL}} & \textbf{Huber loss} & \multirow{2}{*}{\textbf{MSE}} & \multirow{2}{*}{\textbf{MAE}} \\
& & \textbf{(best $\delta$)} & & \\
\hline
A-Indoor               & 5.77 & 6.33 & 7.07 &5.80 \\
A-Pedestrian           & 21.98 & 25.79 & 28.44 & 23.66 \\
A-Mobility             & 13.27 & 15.63 & 17.35 & 13.5\\
A-Outdoor              & 49.48 & 52.74 &  63.44 & 55.16 \\
B-Static-Netflix       & 7.8 & 11.19 & 12.6 &9.34 \\
B-Static-Download      & 11 & 13.49& 13.53 & 12.34\\
B-Static-Amazon\_Prime  & 0.82 & 1.023 & 1.27 & 0.95 \\
B-Driving-Netflix      & 54.1 & 57.6 & 70.13 & 60.27\\
B-Driving-Download     & 36.8 & 38.99 & 45.71 & 38.32 \\
B-Driving-Amazon\_Prime & 17.8 & 18.81 & 19.31 & 18.66 \\
C                      & 21.99 & 22.67 & 26.71& 23.11 \\
\end{tabular}
\end{table}



\begin{figure}[!htb]
\centering
    
\subfigure[Residual-based Adaptive Huber Loss (RAHL) - MAPE=6.55.]{
    \includegraphics[width=0.99\linewidth]{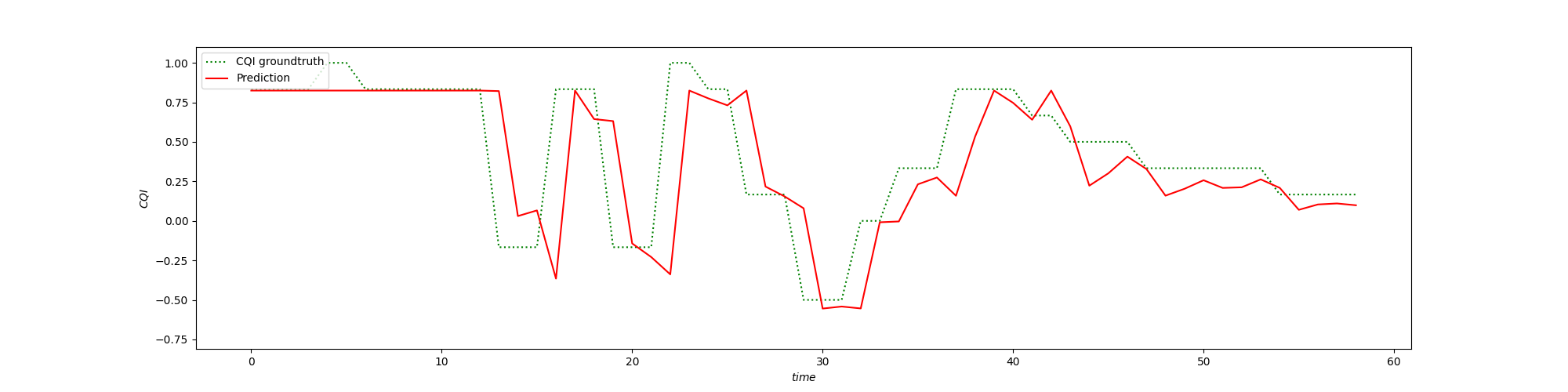}
        \label{fig:(l)}
}
\subfigure[Huber loss (best $\delta$) - MAPE=9.05.]{
    \includegraphics[width=0.99\linewidth]{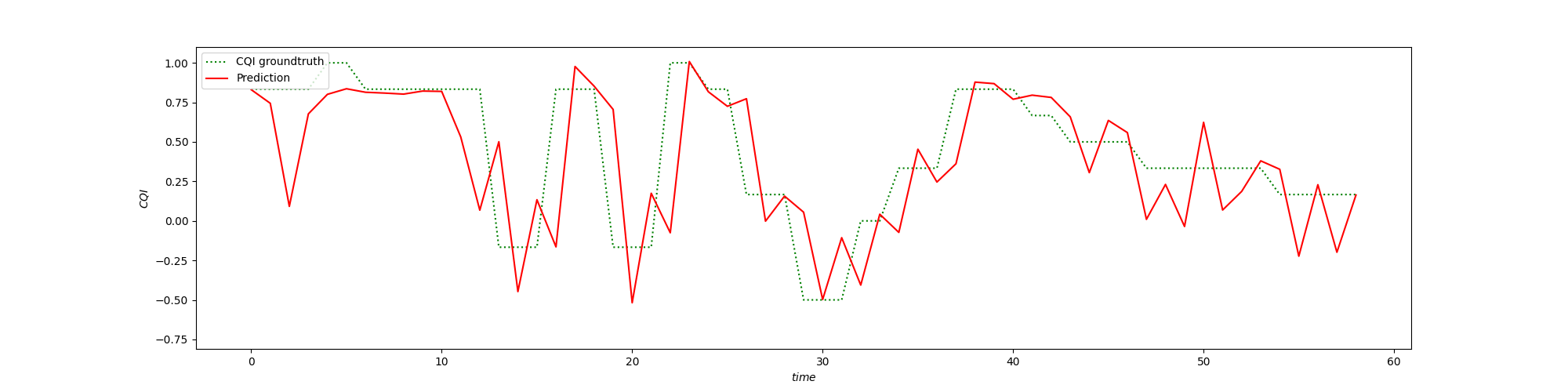}
        \label{fig:(o)}
}
\subfigure[Mean Score Error (MSE) - MAPE=10.37.]{
    \includegraphics[width=0.99\linewidth]{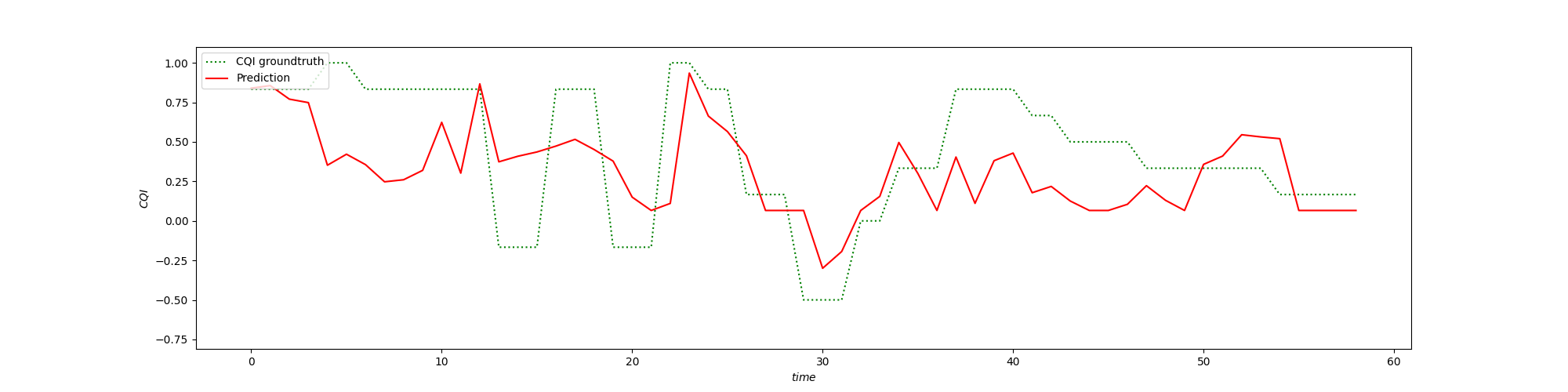}
        \label{fig:(p)}
}
 \subfigure[Mean Absolute Error (MAE) - MAPE=6.73.]{
     \includegraphics[width=0.99\linewidth]{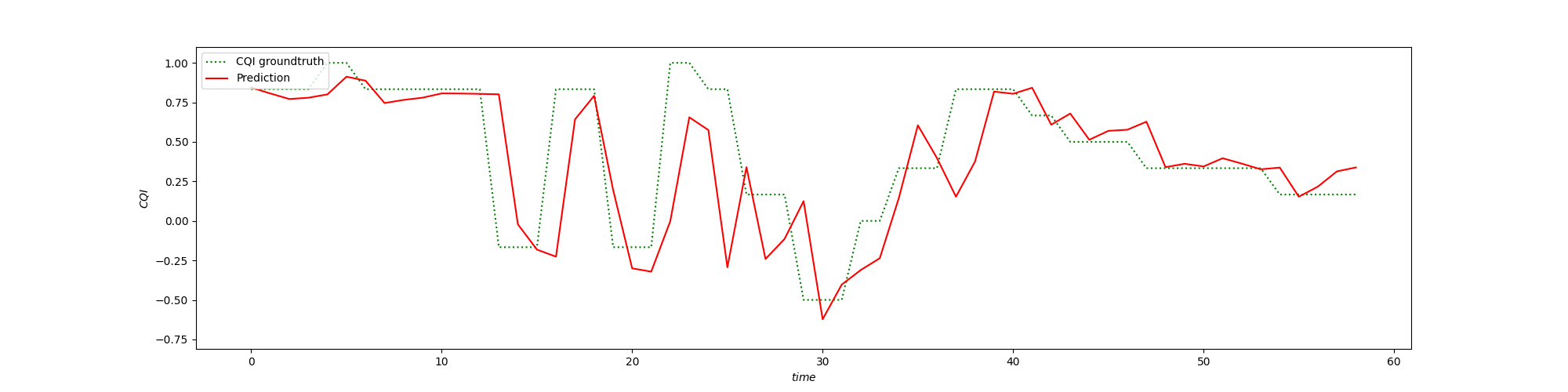}
    \label{fig:(v)}
}

\caption{Comparison of the ground truth value and the model's prediction for a sample from the Mobility category of the dataset A.}
\label{fig:Dataset-Total}
\end{figure}

\begin{figure}[!h]
    \centering
    
    \subfigure[Residual-based Adaptive Huber Loss (RAHL) - MAPE=11.07.]{
        \includegraphics[width=0.99\linewidth]{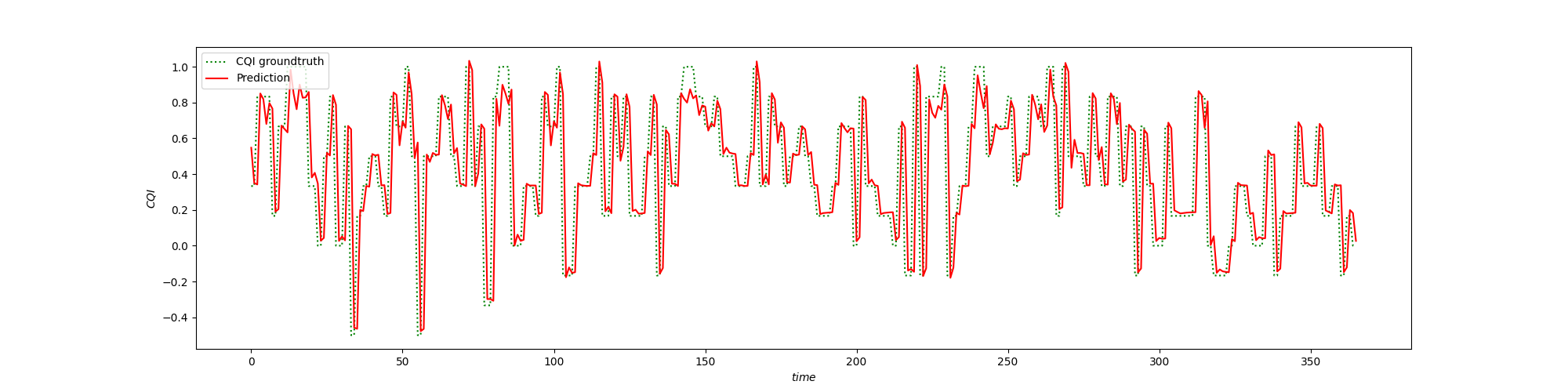}
        \label{fig:(l.)}
    }
    \subfigure[Huber loss (best $\delta$) - MAPE=13.64.]{
        \includegraphics[width=0.99\linewidth]{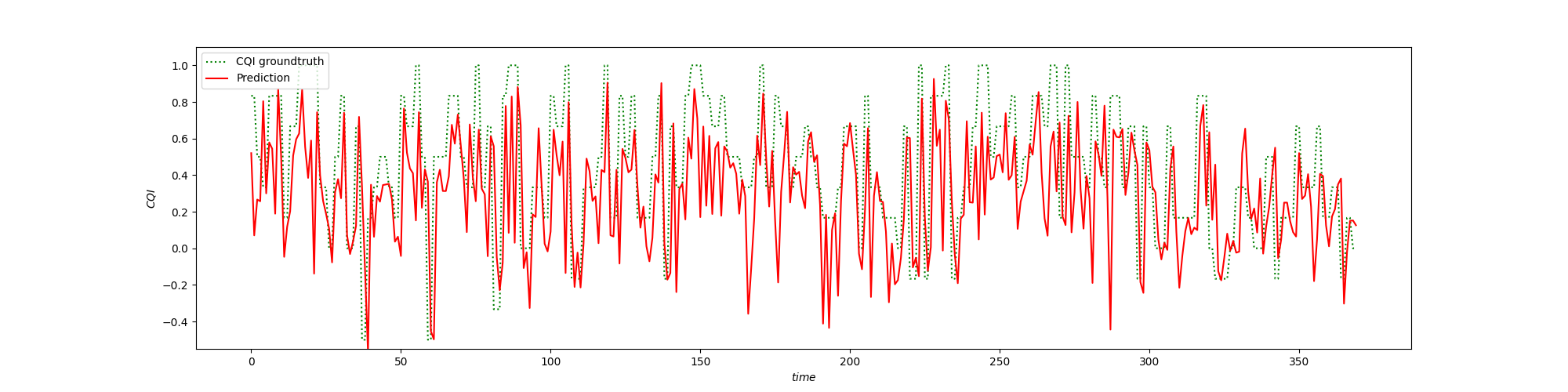}
        \label{fig:(o.)}
    }
    \subfigure[Mean Score Error (MSE) - MAPE=14.20.]{
        \includegraphics[width=0.99\linewidth]{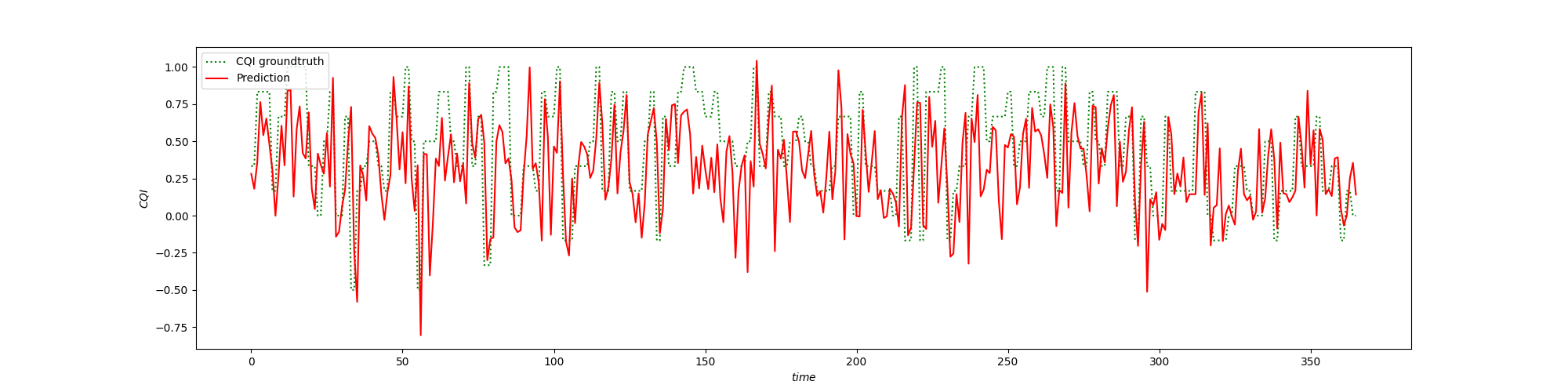}
        \label{fig:(p.)}
    }
     \subfigure[Mean Absolute Error (MAE) - MAPE=12.16.]{
        \includegraphics[width=0.99\linewidth]{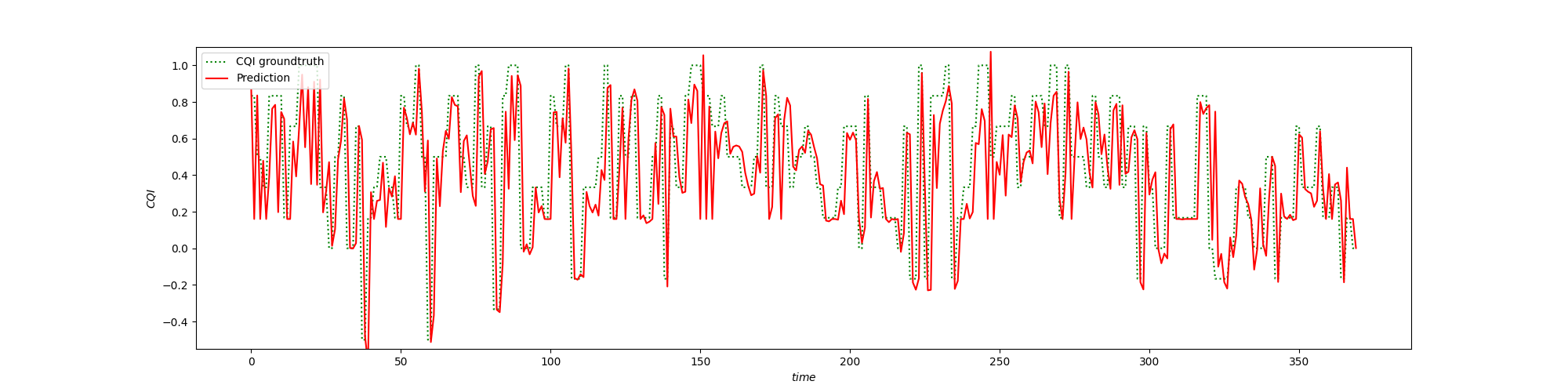}
        \label{fig:(j.)}
    }

    \caption{Comparison of the ground truth value and the model's prediction for a sample from the dataset C.}
    \label{fig:Dataset-Total1}
\end{figure}

To highlight the benefits of performing CQI prediction using a model trained with RAHL, we compare the ground truth value and the model's prediction for some samples from our datasets.
For this analysis, we took samples from two distinct datasets. The first, taken from the Mobility category of the dataset A, comprises a relatively small sample with significant outliers. The second sample, selected from dataset C, was larger, had a more uniform distribution, and exhibited fewer outliers.
Figures~\ref{fig:Dataset-Total}~and~\ref{fig:Dataset-Total1} compare, for each of the chosen samples, respectively, the actual CQI values (green line) and the predictions (red line) made by LSTM models trained with different loss functions.

As the subcaptions of each figure reveal, the MAPE value for RAHL is the lowest in both samples. Additionally, the predictions for the LSTM model trained with RAHL consistently follow the actual CQI values more closely, regardless of sample size (time duration). This indicates the robustness of RAHL against outliers and overall superiority in achieving accurate results for CQI forecasting.

It is of paramount importance to remember that CQI is a metric for quantifying the quality of the radio channel between the UE and the base station. The CQI enables the base station to dynamically adapt the modulation for each UE so that the data rate can be optimized. As a consequence, making a wrong CQI prediction will affect negatively how the modulation is configured and how the resource allocation in the 5G network will be done. As an example, by analyzing the predictions made by the LSTM model trained with Huber loss (best $\delta$) in Figure~\ref{fig:(o)}, we can observe from time 60 that the prediction is (wrongly) going up and down, suggesting some network issue. However, the actual CQI values indicate that the network connection is stable for most of the time. The same can be observed in Figure \ref{fig:(p)} in which the CQI prediction is wrong during almost the entire period.

\begin{figure}[!h]
    \centering
    
    \subfigure[Dataset A-Mobility.]{
        \includegraphics[width=0.45\linewidth]{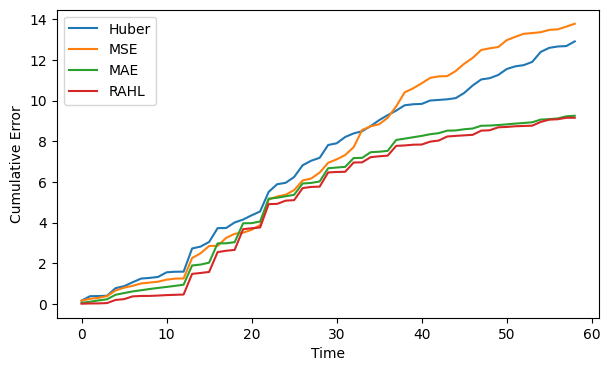}
        \label{fig:(uu)}
    }
    \subfigure[Dataset C.]{
        \includegraphics[width=0.47\linewidth]{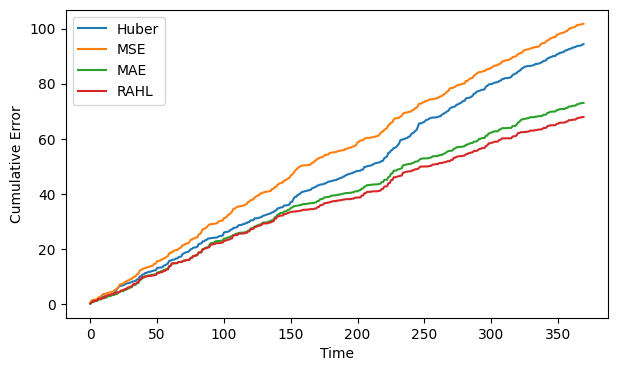}
        \label{fig:(ll)}
    }

    \caption{Absolute percentage error accumulated over time for different losses.}
    \label{fig:Cumulative}
\end{figure}

Figure~\ref{fig:Cumulative} presents a different view of such results, showing how the absolute percentage error is accumulated over time. In this way, we can analyze the error patterns for LSTM models trained with different loss functions. As expected, the cumulative error associated with RAHL grows slower than that of other loss functions and, for this reason, it is demonstrably more effective for CQI prediction.

\section{Conclusion} \label{sec:Conclusion}

CQI is the most important metric to represent the quality of the 5G channel. It is used by the base station to make resource allocation, modulation and coding. The channel quality directly affects the data rate and the usage of the network capacity, which at the end, will affect the QoE of the user. When using ML models for CQI prediction, the lower the error (MAPE) the better the user experience. 

Our paper introduces an advancement in minimizing error rates through the introduction of the RAHL method. The experiments showcased in Figures \ref{fig:Dataset-Total}, \ref{fig:Dataset-Total1} and \ref{fig:Cumulative} demonstrate the superior performance of RAHL compared to other alternatives, such as Huber loss, MSE, and MAE, in the context of CQI prediction. Our findings have significant implications for real-world applications, where accurate predictions are crucial. The elimination of manual hyperparameter tuning in RAHL addresses challenges associated with subjectivity, data-dependent considerations, computational costs, potential overfitting, and the delicate balance between accuracy and robustness. While our study has provided valuable insights, we acknowledge its limitations, and future work could explore the applicability of RAHL across diverse datasets and neural network architectures. Additionally, considering the dynamic nature of loss functions, our work opens avenues for further research into adaptive approaches that can enhance performance across various tasks. Our work offers practical recommendations that can be valuable for both practitioners and researchers. The innovative aspects of RAHL not only enhance prediction accuracy but also simplify the process of selecting suitable hyperparameters, making it a valuable addition to the machine learning toolkit. In summary, our proposed RAHL offers a versatile and effective solution for error rate minimization.

\section*{Acknowledgments}
This study was financed in part by the Coordenação de
Aperfeiçoamento de Pessoal de Nível Superior - Brasil (CAPES) - Finance Code 001.

\bibliographystyle{sbc}
\bibliography{sbctemplate}

\end{document}